\documentclass[aps,prc,twocolumn,preprintnumbers,secnumarabic,amsmath,amssymb,nofootinbib,floatfix,superscriptaddress]{revtex4}
\usepackage{graphicx}  
\usepackage{dcolumn}   
\usepackage{bm}        
\usepackage{amssymb}   
\usepackage{amsmath}
\usepackage{slashed}
\usepackage{hyperref}
\usepackage{color}
\usepackage{epsfig}
\usepackage{comment}
\usepackage{epstopdf}
\usepackage[utf8]{inputenc}
\usepackage[T1]{fontenc}
\usepackage{caption}  
\usepackage{subcaption} 
\newcommand{\gsim}{\buildrel > \over {_\sim}}
\renewcommand{\raggedright}{\leftskip=0pt \rightskip=0pt plus 0cm}

\begin{document}

\preprint{ACFI-T16-14}x

\title{{\bf Right-Handed Neutrinos and T-Violating, P-Conserving Interactions}}

\author{Basem Kamal El-Menoufi}
\affiliation{Amherst Center for Fundamental Interactions, Department of Physics,
University of Massachusetts\\
Amherst, MA  01003, USA}
\author{Michael J. Ramsey-Musolf}
\affiliation{Amherst Center for Fundamental Interactions, Department of Physics,
University of Massachusetts\\
Amherst, MA  01003, USA}
\affiliation{Kellogg Radiation Laboratory, California Institute of Technology,
Pasadena, CA 91125 USA}
\author{Chien-Yeah Seng}
\affiliation{Amherst Center for Fundamental Interactions, Department of Physics,
University of Massachusetts\\
Amherst, MA  01003, USA}

\begin{abstract}
We show that experimental probes of the P-conserving, T-violating
triple correlation in polarized neutron or nuclear  $\beta$-decay
provide a unique probe of possible T-violation at the TeV scale in
the presence of right-handed neutrinos. In contrast to other
possible sources of semileptonic T-violation involving only
left-handed neutrinos, those involving right-handed neutrinos are
relatively unconstrained by present limits on the permanent electric
dipole moments of the electron, neutral atoms, and the neutron. On
the other hand, LHC results for $pp\to e+$ missing transverse energy
imply that an order of magnitude of improvement in $D$-coefficient
sensitivity would be needed for discovery.
Finally, we discuss the interplay with the scale of neutrino mass
and naturalness considerations.

\end{abstract}

\maketitle

\section{Introduction}

The search for time-reversal violation (TV) has long been a subject
of considerable experimental and theoretical interest. It is
partially motivated by the need for CP-violation beyond that encoded
in the Standard Model (SM) Cabibbo-Kobayashi-Maskawa (CKM) matrix to
explain the cosmic baryon asymmetry\cite{Sakharov:1967dj}. Assuming
CPT is a good symmetry of nature, searches for TV provide a probe of
this possible CP-violation. Experimentally, the parity (P)- and
T-violating (PVTV) sector is being probed with great sensitivity
through electric dipole moment (EDM) searches, with the three most
stringent limits having been obtained for the $^{199}$Hg
atom\cite{Graner:2016ses}, the electron (extracted from the ThO
molecule)\cite{Baron:2013eja}, and the
neutron\cite{Afach:2015sja,Baker:2006ts}.
hand, the P-conserving and T-violating (PCTV) sector ( equivalent to
C- and CP-violation assuming CPT) has received considerably less
attention. Experimental efforts in the sector include measurement of
anomalous $\eta$-decay channels such as $\eta\rightarrow
2\pi^0\gamma$, $3\pi^0\gamma$, $3\gamma$ \cite{Beringer:1900zz} and
the $D$-coefficient in the $\beta$-decay of polarized neutrons
\cite{Mumm:2011nd} and $^{19}$Ne \cite{Hallin:1984mr}. These
processes are sensitive probes of \lq\lq new physics\rq\rq\, because
the Standard Model (SM) contributions are usually small
\cite{Dicus:1975cz,Herczeg:1997se}. There is a SM final-state
interaction that could mimic a non-zero $D$-coefficient in
$\beta$-decay at order $10^{-5}$ for neutron \cite{Callan:1967zz}
and $10^{-4}$ for $^{19}$Ne \cite{Hallin:1984mr} but the application
of heavy baryon effective field theory allows a precise computation
of this contribution (up to 1\% accuracy in the case of neutron
\cite{Ando:2009jk}).

Theoretically, the effect of PCTV physics due to beyond Standard
Model (BSM) interactions  can be studied in a model-independent way
using effective field theory (EFT). In this approach, one has
integrated out the BSM heavy degrees of freedom (DOF). In this
context, it was observed in Ref. \cite{Conti:1992xn} that any EDM
limits imply severe bounds on PCTV observables since a PCTV
interaction in the presence of P-violating SM radiative corrections
will induce an EDM. While special exceptions to this argument may
occur \cite{RamseyMusolf:1999nk,Kurylov:2000ub}, the question
remains as to the prospective impact of,  and motivation for,
improved probes of flavor-conserving PCTV observables. Recently, the
authors of Ref. \cite{Ng:2011ui} addressed this question in the EFT
context, studying the contribution of the \lq\lq left-right four
fermion\rq\rq\, (LR4F) operator to the $D$-coefficient of the
neutron $\beta$-decay (defined below). They find that the neutron
EDM sets an indirect bound on the $D$-coefficient that is three
orders of magnitude more stringent than its direct experimental
bound.


In this paper, we observe that there exists a set of dimension-six
four-fermion operators involving right-handed neutrinos that (a)
contribute to the $D$-coefficient and (b) are relatively
unconstrained by EDM limits. Because the SM charge changing weak
interaction involves purely left-handed leptons, this contribution
to neutron decay does not interfere linearly with the SM
contribution, resulting in a quadratic, rather than linear,
dependence on the operator Wilson coefficients. Nonetheless, present
limits on $D$ probe the TeV mass scale. We also show that while a
subset of these operators  generate hadronic EDMs, their effects are
suppressed by loop factors as well as $\Lambda_{\chi}/v$ where
$\Lambda_{\chi}\sim 1$ GeV is the chiral symmetry breaking scale and
$v=246$ GeV is the Higgs vacuum expectation value (VEV). The
resulting neutron EDM sensitivity to $\Lambda$ is also at the TeV
scale and does not preclude a non-zero result in a next generation
$D$-coefficient probe.

Interestingly, indirect constraints from T-conserving observables
may be more severe. These observables include Large Hadron Collider
(LHC) results for the process $pp\to e + X + MET$ (missing
transverse energy) and neutrino mass. The latter constraints also
rely on naturalness considerations, a somewhat subjective criteria.
The former imply that an order of magnitude improvement in
$D$-coefficient sensitivity would be required in order to discover
evidence for PCTV right-handed neutrino interactions.

Our analysis leading to these conclusions is organized as follows.
In Sec. \ref{sec:vR} we introduce the relevant set of dimension-6
operators and discuss the experimental $D$-coefficient constraint on
their Wilson coefficients. We then compare this constraint to those
implied by LHC data, hadronic EDMs as well as neutrino mass and
naturalness considerations. For comparison, we perform in Sec.
\ref{sec:novR} a similar analysis of other dimension-6 operators
that do not involve right-handed neutrinos. We show that any attempt
to evade current EDM constraints and yet keep the size of the
$D$-coefficient experimentally accessible would involve fine tuning
at the $10^{-11}$ level. We conclude in Sec. \ref{sec:conclusion}.

\section{\label{sec:vR}Dimension-six operators with right-handed neutrinos}

The PCTV observable of interest in $\beta$-decay involves a triple
correlation of the spin of the decaying particle and the momenta of
the outgoing leptons that enters the differential $\beta$-decay
rate. In what follows, we focus on neutron, for which the
experimental bound on the PCTV triple correlation is the most
stringent. However, the discussion below can be easily generalized
to other cases. The differential decay rate for a polarized neutron
is given by:
\begin{eqnarray}
\nonumber
\frac{d\Gamma}{dE_ed\Omega_ed\Omega_\nu}=\frac{G_F^2V_\mathrm{ud}^2}{(2\pi)^5}(g_V^2+3g_A^2)|\vec{p}_e|E_eE_\nu^2\\
\times
\left[1+a\frac{\vec{p}_e\cdot\vec{p}_\nu}{E_eE_\nu}+\hat{s}\cdot(A\frac{\vec{p}_e}{E_e}+B\frac{\vec{p}_\nu}{E_\nu}+D\frac{\vec{p}_e\times\vec{p}_\nu}{E_eE_\nu})\right]
\end{eqnarray}
where $\hat{s}$ is the unit polarization vector of the neutron;
$\vec{p}_e$ and $\vec{p}_\nu$  are the electron and anti-neutrino
momenta, respectively, with corresponding energies $E_{e(\nu)}$;
$G_F$ is the Fermi constant; and $V_\mathrm{ud}$ is the first
generation element of the Cabibbo-Kobayashi-Maskawa (CKM) matrix.
The most stringent experimental limit on the $D$-coefficient is
given by $D=(-0.96\pm1.89\pm1.01)\times10^{-4}$ \cite{Mumm:2011nd}
which translates into an upper bound of $|D|<4\times 10^{-4}$ at
90\% CL \cite{Vos:2015eba}.

Theoretically, a non-vanishing contribution can be generated by the
interference of amplitudes involving a small set of dimension $d=6$
effective operators. Considering only first generation SM fermions
and requiring SM gauge invariance, one finds a limited set of such
$d=6$ TV operators (see Ref. \cite{Grzadkowski:2010es} for a
complete list of gauge-invariant $d=6$ operators involving SM
fields). As we discuss in Section \ref{sec:dipole}, EDM constraints
imply severe bounds on the contribution of these operators to $D$.
Extending the set of fields to include right-handed (RH) neutrinos,
one finds an additional set of four-fermion operators that
contribute to $D$ at tree-level and that are relatively immune to
EDM constraints \cite{Cirigliano:2012ab}:


\begin{eqnarray}\label{eq:Oi}
\hat{O}_1&=&\frac{c_1(\mu)}{\Lambda^2}\bar{L}^i\nu_R\bar{u}_RQ^i+h.c.\nonumber\\
\hat{O}_2&=&\frac{c_2(\mu)}{\Lambda^2}\varepsilon^{ij}\bar{L}^i\nu_R\bar{Q}^jd_R+h.c.\nonumber\\
\hat{O}_3&=&\frac{c_3(\mu)}{\Lambda^2}\varepsilon^{ij}\bar{L}^i\sigma^{\mu\nu}\nu_R\bar{Q}^j\sigma_{\mu\nu}d_R+h.c.
\end{eqnarray}
where $\Lambda$ is the BSM mass scale and $\mu$ is the
renormalization scale. These operators are analogous to the
semi-leptonic four-fermion operators of type $(\bar{L}R)(\bar{L}R)$
and $(\bar{L}R)(\bar{R}L)$ in Ref. \cite{Grzadkowski:2010es}. Also
notice that the Wilson coefficients $c_1-c_3$ are functions of the
renormalization scale $\mu$, which as to be taken as the hadronic
scale when we discuss the bounds of the Wilson coefficients from
low-energy experiments.

It is straightforward to compute the contributions of
$\hat{O}_{1-3}$ to the $D$-coefficient. The dominant affect is
quadratic in the $c_i/\Lambda^2$, as the linear interference term is
suppressed by the neutrino mass.
Following Ref.~\cite{Jackson:1957zz}, we obtain, to leading
non-trivial order in $\{c_i\}$,
\begin{equation}
\label{eq:Dcoeff}
D=-\frac{g_Sg_T}{\Lambda^4}\frac{1}{G_F^2V_{ud}^2(g_V^2+3g_A^2)}\mathrm{Im}[(c_1-c_2)c_3^*]\Big\vert_{\mu=\mu_h}
\end{equation}
where $g_S$ and $g_T$ are the nucleon scalar and tensor charges,
respectively, and $\mu_h\approx1$ GeV is the hadronic scale.

Even though one pays a price in BSM sensitivity owing to a quadratic
rather than linear dependence on the $c_i/\Lambda^2$, the gain
achieved by avoiding EDM constraints is considerable (see Section
\ref{sec:dipole}). Taking the updated lattice calculation of
$g_S=0.97(12)(6)$ and $g_T=0.987(51)(20)$
\cite{Bhattacharya:2016zcn}, we obtain:
\begin{equation}\label{eq:bound}
|\frac{\mathrm{Im}[(c_1-c_2)c_3^*]}{\Lambda^4}|\Big\vert_{\mu=\mu_h}<3\times10^{-1}\
\mathrm{TeV}^{-4}
\end{equation}
If we take $\{c_i\}\sim c$ without distinguishing the real and
imaginary part, then this inequality implies that existing
$D$-coefficient studies  probe BSM T-violating interactions with RH
neutrinos with a sensitivity of $(v/\Lambda)^2c\sim3\times 10^{-2}$
at $\mu=\mu_h$. One could estimate the sensitivity to $\Lambda$ by
assuming that $c_i\sim1$ at $\mu\approx\Lambda$. QCD running in
$\overline{\mathrm{MS}}$ scheme gives
$c_{1,2}(\Lambda)\approx0.56c_{1,2}(\mu_h)$ and $c_3(\Lambda)\approx1.2c_3(\mu_h)$ for $\Lambda>m_W$ where $m_W$ is the mass of the W-boson (see,
e.g. Ref. \cite{Broadhurst:1994se}). Then, the current bound implies $\Lambda\gsim 1$
TeV.

\begin{figure}
\centering
\begin{subfigure}[b]{0.25\textwidth}
\includegraphics[width=\textwidth]{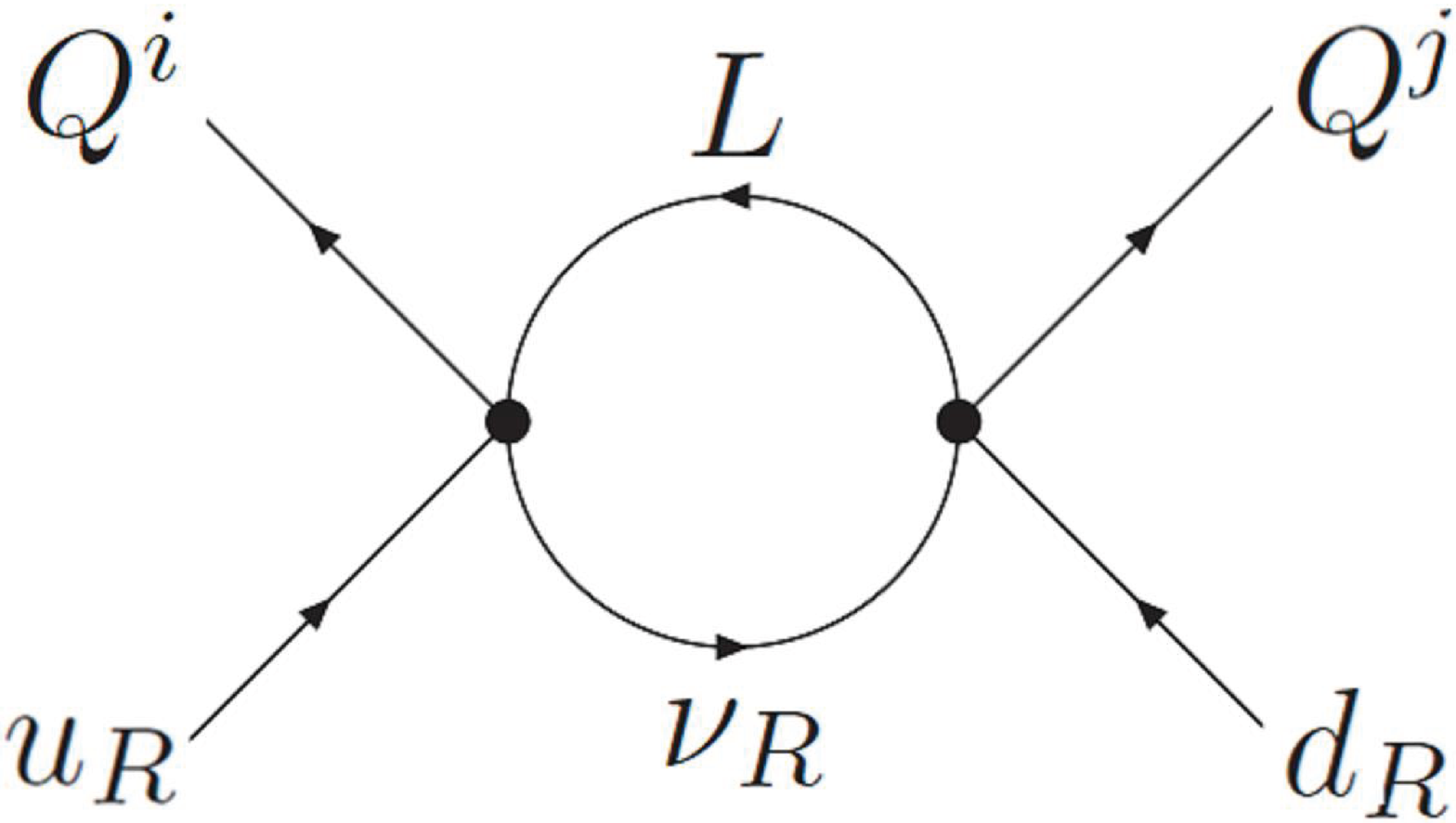}
\caption{\label{fig:4q}(a)}
\end{subfigure}
\begin{subfigure}[b]{0.25\textwidth}
\includegraphics[width=\textwidth]{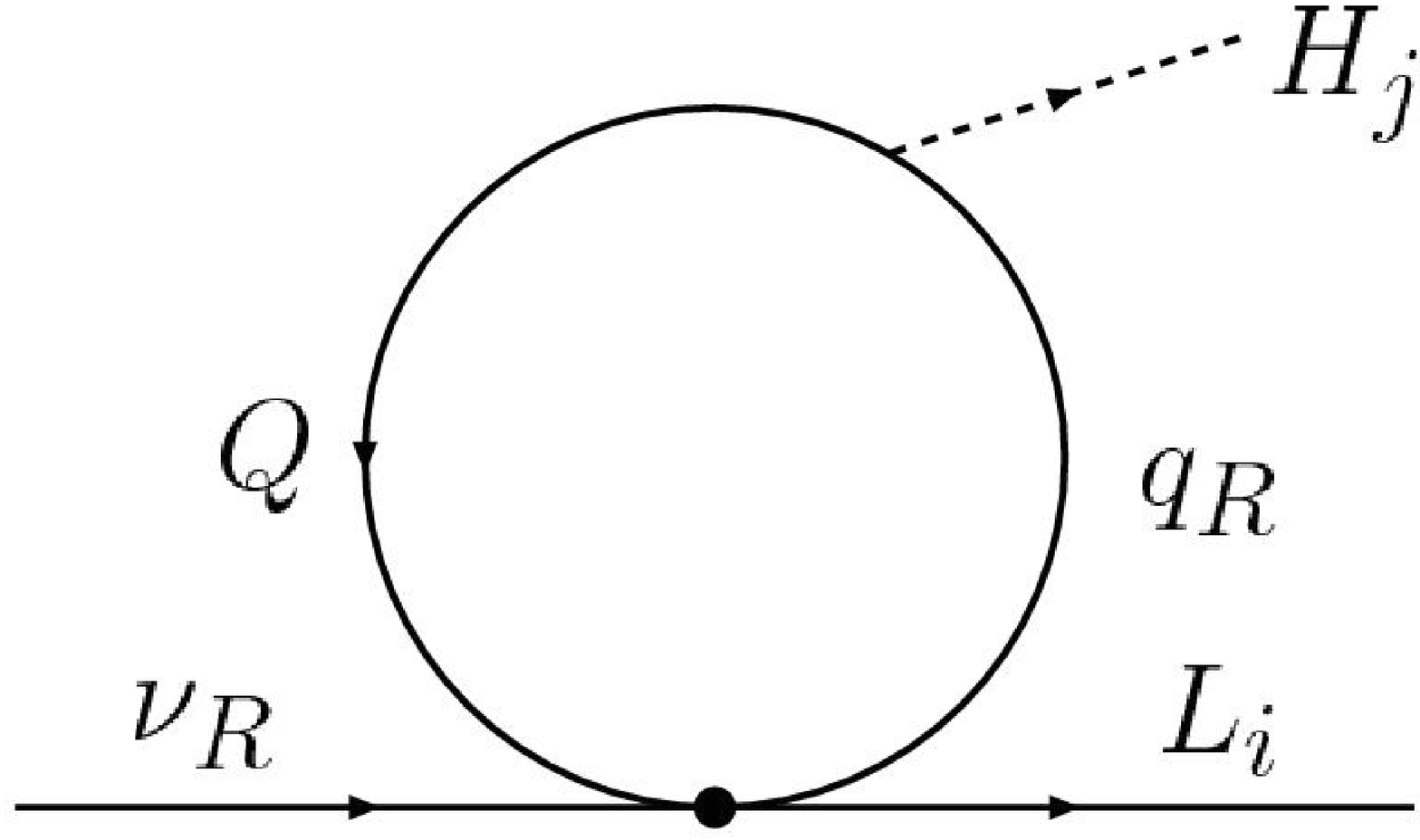}
\caption{\label{fig:numass}(b)}
\end{subfigure}
\caption{\raggedright Leading loop contributions that provide
indirect bounds on $c_1$ and $c_2$. Figure (a) induces a four-quark
operator that generates hadronic EDMs. Figure (b) generates a
neutrino mass after electroweak symmetry breaking.}
\end{figure}

The operators ${\hat O}_{1-3}$ can induce hadronic EDMs at one-loop order, but their contributions also scale
quadratically with the $c_i/\Lambda^2$.
In particular, the combination of $\hat{O}_1$ and $\hat{O}_2$ may
induce the CP-odd four-quark operator \cite{Engel:2013lsa}
\begin{equation}
\frac{C_{quqd}^{(1)}(\mu)}{\Lambda^2}
\varepsilon^{ij}\bar{Q}^iu_R\bar{Q}^jd_R+h.c.
\end{equation}
via the one-loop graph of Fig. \ref{fig:4q}. Contributions from loop
momenta $k < \Lambda$ vanish, as seen explicitly in dimensional
regularization (DR), because the amplitude involves a
quadratically-divergent integral with massless propagators and
because it is infrared finite. Non-vanishing contributions result
from $k\gsim \Lambda$ that are associated with matching onto the
{\em a priori} unknown ultraviolet complete theory that generates
the non-vanishing $c_i$. Estimating these matching contributions
using a cut-off regulator \cite{Erwin:2006uc} yields\footnote{It is
possible that in the full theory, a symmetry implies vanishing
matching contributions, but we will be more general here.}
\begin{equation}\label{eq:4fto4q}
\frac{C_{quqd}^{(1)}(\Lambda)}{\Lambda^2}\sim\frac{\Lambda^2}{16\pi^2}\frac{c_1^*c_2}{\Lambda^4}\Big\vert_{\mu=\Lambda}=\frac{c_1^*c_2}{16\pi^2\Lambda^2}\Big\vert_{\mu=\Lambda}
\end{equation}

This four-quark operator will in turn induce a neutron EDM $d_n$. To
evaluate this contribution, one must first evolve $C_{quqd}^{(1)}$
from $\mu=\Lambda$ down to $\mu=\mu_h$. In principle, this can only
be done if one knows the exact value of $\Lambda$. However, since
the evolution depends only logarithmically on $\Lambda$, it is
reasonable to take $\Lambda\sim1$ TeV  as an illustration, giving
$C_{quqd}^{(1)}(\mu_h)=7.2C_{quqd}^{(1)}(\Lambda)$
\cite{Dekens:2013zca}. 

Next, in order to find the relation between $C_{quqd}^{(1)}(\mu_h)$
and the induced hadronic EDMs one needs to compute corresponding
hadronic matrix elements. First-principle calculations of such
matrix elements are challenging, and presently only exist for simple
systems such as $\rho$-meson (see, {\em e.g.} Ref.
\cite{Pitschmann:2012by} and references therein). The results of
such calculations are generally consistent with the
order-of-magnitude estimation based on na{\"i}ve dimensional
analysis (NDA) \cite{Manohar:1983md,Weinberg:1989dx,deVries:2012ab},
so here we shall also provide an NDA estimation of $d_n$:
\begin{eqnarray}\label{eq:NDA} d_n&\sim&
e\frac{\Lambda_\chi}{16\pi^2}\frac{\mathrm{Im}C_{quqd}^{(1)}(\mu_h)}{\Lambda^2}\nonumber\\
&\approx&
e\frac{\Lambda_\chi}{16\pi^2}\times7.2\frac{\mathrm{Im}C_{quqd}^{(1)}(\Lambda)}{\Lambda^2}\nonumber\\
&\approx& 9.4\times
10^{-23}(\frac{v}{\Lambda})^2\mathrm{Im}\{c_1^*c_2\}|_{\mu=\Lambda}e\:\mathrm{cm}.
\end{eqnarray}
This EDM is suppressed by
$1/(16\pi^2)^2$ as well as $\Lambda_\chi/v$. Given the current upper
bound $d_n<3.0\times10^{-26}e\:\mathrm{cm}$ at 90\% CL
\cite{Afach:2015sja} we see that the existing neutron EDM limits are
probing $(v/\Lambda)^2c^2\sim3\times10^{-4}$ at $\mu=\Lambda$ which
implies $\Lambda\gsim 10$TeV if $c(\Lambda)\sim1$.

At first glance, the neutron EDM sensitivity to  $\Lambda$ is slightly tighter than that of the
$D$-coefficient. However, since both estimations made in Eq.
\eqref{eq:4fto4q} and \eqref{eq:NDA} allow an error within an order
of magnitude, one may reasonably conclude that the sensitivities of
$d_n$ and the $D$-coefficient are
comparable. Furthermore, hadronic and atomic EDMs depend only on
$c_1^*c_2$ and provide no direct constraint on the contribution from
$c_jc_3^\ast$ ($j=1,2$) in Eq.~(\ref{eq:Dcoeff}).

We now consider constraints from T conserving observables. First, we
note that LHC studies of the process $pp\rightarrow e+X+MET$ place
stringent bounds on the operators in \eqref{eq:Oi}\footnote{We thank
M. Gonzales-Alonso for pointing out these constraints.}.
Following Ref. \cite{Cirigliano:2012ab}, one may define two
dimensionless quantities:
\begin{eqnarray}
\tilde{\epsilon}_S&=&-\frac{c_1-c_2}{2\sqrt{2}G_F
V_{ud}\Lambda^2}\nonumber\\
\tilde{\epsilon}_T&=&\frac{c_3}{2\sqrt{2}G_F V_{ud}\Lambda^2}.
\end{eqnarray} The contribution from $\hat{O}_{1,2,3}$ to the total cross-section $\sigma_{tot}$ of the $pp\rightarrow e+X+MET$ process measured by LHC can be written as $\sigma_{tot}=\sigma_S|\tilde{\epsilon}_S|^2+\sigma_T|\tilde{\epsilon}_T|^2$. Therefore LHC is sensitive to $|\tilde{\epsilon}_S|$ and $|\tilde{\epsilon}_T|$
while the
$D$-coefficient probes the combination of products
$\mathrm{Re}\tilde{\epsilon}_T\mathrm{Im}\tilde{\epsilon}_S-\mathrm{Re}\tilde{\epsilon}_S\mathrm{Im}\tilde{\epsilon}_T$.
The bounds on the ${\tilde\epsilon}$ parameters obtained in Ref.
\cite{Cirigliano:2012ab} assume contributions from one operator at a
time. However, when comparing with the $D$-coefficient sensitivity,
one must take both $\tilde{\epsilon}_S$ and $\tilde{\epsilon}_T$,
since the $D$-coefficient probes products of the two. Recasting the
analysis of Ref.~\cite{Cirigliano:2012ab} is nevertheless
straightforward because $\sigma_S$ and $\sigma_T$ are known. The
constraint equation in Ref.~\cite{Cirigliano:2012ab} then implies an
elliptical bound in the $|\tilde{\epsilon}_S|-|\tilde{\epsilon}_T|$
plane. One should also remember that the LHC constraints should be
run down to $\mu=\mu_h$ for a fair comparison with the
$D$-coefficient.

\begin{figure}
\centering
\includegraphics[scale=0.30]{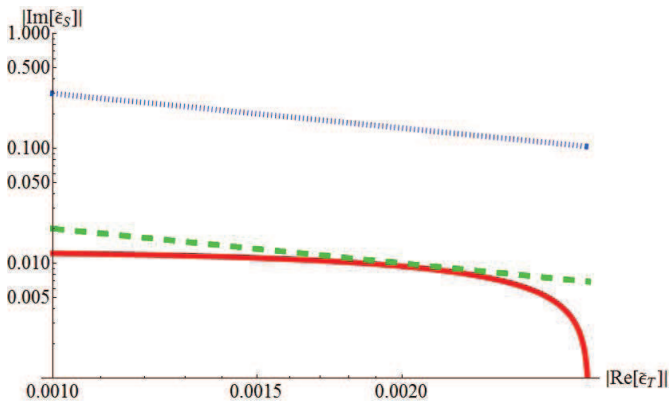}
\caption{\raggedright\label{fig:LHCvsD}(Color online)
Exclusion plot for $\mathrm{Re}\tilde{\epsilon}_{T}$ and
$\mathrm{Im}\tilde{\epsilon}_{S}$ at $\mu=\mu_h$ from the 7-TeV LHC
data (red solid line) as well as the bound from the neutron
$D$-coefficient with the current precision level (blue dotted line) and 15 times of the current precision level (green dashed line) respectively assuming
$\mathrm{Re}\tilde{\epsilon}_{S}=\mathrm{\mathrm{Im}\tilde{\epsilon}_{T}=0}.$}
\end{figure}

Since there are four real parameters in the problem (the Re and Im
parts of $\tilde\epsilon_{S,T}$) , it us useful to make simplifying
assumptions in order to compare the LHC and $D$-coefficient
sensitivities. To that end, we will
assume for the moment that
$\mathrm{Re}\tilde{\epsilon}_S=\mathrm{Im}\tilde{\epsilon}_T=0$ so
both the LHC  and the neutron $D$-coefficient results set
constraints on $\mathrm{Re}\tilde{\epsilon}_T$ and
$\mathrm{Im}\tilde{\epsilon}_S$ (see Fig. \ref{fig:LHCvsD}). In this
case, one sees that the sensitivity of neutron decay to the
$D$-coefficient has to be improved by roughly a factor of 15 in
order to match the sensitivity of the  7-TeV  LHC results. Results
at $\sqrt{s}=8$ TeV for the same channel at are also available. As
there is no significant deviation from SM prediction
\cite{ATLAS:2014pna,Khachatryan:2014tva}, the LHC bound on
$|\tilde{\epsilon}_S|$ and $|\tilde{\epsilon}_T|$ will be even more
stringent than quoted above,  although a detailed analysis has yet
to be performed\footnote{The LHC sensitivity will, of course,
improve further with the data obtained from Run II .}.


One may also derive interesting but less direct constraints on  $\hat{O}_1$ and $\hat{O}_2$ from the scale of neutrino mass and naturalness considerations.
Above the electroweak scale, the leading contribution to $m_\nu$
comes from a one-loop diagram with a quark Yukawa insertion,
inducing the Yukawa interaction term $\bar{L}\tilde{H}v_R$, as shown
in Fig. \ref{fig:numass}. Again this contribution vanishes in DR so
we estimate it using simple dimensional analysis. After electroweak
symmetry breaking, one obtains
\begin{equation}
m_\nu\sim\frac{c_i}{\Lambda^2}\frac{\Lambda^2}{16\pi^2}m_q=\frac{c_i
m_q}{16\pi^2}
\end{equation}
where $m_q$ is the light quark mass and $i=1,2$. Taking $m_\nu<1$ eV
and $m_q\approx 5$ MeV we obtain $c_i<3\times10^{-5}$. We
stress that this bound is not  airtight, as the result may
vary considerably, depending on the specific symmetry of the underlying BSM
scenario. Neutrino mass naturalness bounds also do not constrain the tensor
interaction strength $c_3$. Should a next generate $D$-coefficient measurement yield a non-vanishing result, the comparison with
neutrino mass naturalness considerations would provide interesting input for model-building.


\section{\label{sec:novR}Operators without right-handed neutrinos}
\label{sec:dipole}

In contrast to the discussion of Section \ref{sec:vR}, we consider here $d=6$ operators that contain only
left-handed (LH) neutrino fields and show that any contributions to the $D$-coefficient are severely constrained by present EDM limits.
In the four-fermion sector, the only operator that gives a
tree-level $D$-coefficient scaling linearly with the BSM coupling
strength has the form of $\bar{u}_R\gamma^\mu d_R\bar{e}_L\gamma_\mu
\nu_L$ as discussed in Ref. \cite{Ng:2011ui}. It is actually derived
from a gauge-invariant dim-6 operator:
\begin{equation}
\label{eq:hud}
\hat{O}_{Hud}=i\frac{C_{Hud}}{\Lambda^2}(\tilde{H}^\dagger
D_\mu H)(\bar{u}_R\gamma^\mu d_R)+h.c..
\end{equation}
Below the electroweak scale, exchange of the $W$-boson contained in
the covariant derivative with the left-handed charged weak current
leads to both the semi-leptonic four-fermion operator listed above
as well as
a four-quark operator of the form
$\bar{u}_R\gamma^\mu d_R\bar{d}_L\gamma_\mu u_L$. Both operators share the
same Wilson coefficient (up to $V_{ud}$), which
 is tightly constrained by the the four-quark contribution to the neutron EDM.

The three remaining semi-leptonic four-fermion operators that
contain T-odd components are the scalar and tensor operators of the
type $(\bar{L}R)(\bar{L}R)$ and $(\bar{L}R)(\bar{R}L)$
\cite{Grzadkowski:2010es,Engel:2013lsa}:
\begin{eqnarray}\label{eq:4fermion}
\hat{O}_{ledq}&=&i\frac{\mathrm{Im}C_{ledq}}{\Lambda^2}\bar{L}^ie_R\bar{d}_RQ^i+h.c.\nonumber\\
\hat{O}_{lequ}^{(1)}&=&i\frac{\mathrm{Im}C_{lequ}^{(1)}}{\Lambda^2}\varepsilon^{ij}\bar{L}^ie_R\bar{Q}^ju_R+h.c.\nonumber\\
\hat{O}_{lequ}^{(3)}&=&i\frac{\mathrm{Im}C_{lequ}^{(3)}}{\Lambda^2}\varepsilon^{ij}\bar{L}^i\sigma^{\mu\nu}e_R\bar{Q}^j\sigma_{\mu\nu}u_R+h.c.
\end{eqnarray}
Similar to the operators in Eq. \eqref{eq:Oi}, they induce a
$D$-coefficient that scales quadratically with $c_i/\Lambda^2$.
However, $\hat{O}_{ledq}$, $\hat{O}_{lequ}^{(1)}$,
$\hat{O}_{lequ}^{(3)}$ contribute linearly to EDMs of
paramagnetic atom and molecules and diamagnetic atoms at tree level
as well as hadronic and electron EDMs at the one-loop level.
In particular, the ACME limit on the EDM of ThO molecule
implies a strong constraint on
$\mathrm{Im}(C_{ledq}-C_{ledq}^{(1)})$ (see, e.g. Ref.
\cite{Chupp:2014gka}). The resulting indirect constraints on the
associated $D$-coefficient contributions are severe.


The remaining class of operators that give rise to the
$D$-coefficient at tree-level are dipole-like operators. One may
wonder whether EDM constraints to such operators may be avoided with
an appropriate choice of Wilson coefficients at low energy. We will
show, however, that this is not possible without fine-tuning at the
level of many orders of magnitude.
To simplify our discussion, let us concentrate on the dipole-like
operators in the purely leptonic sector:
\begin{eqnarray}\label{eq:op}
\hat{O}_{eB}&=&i\frac{g'\mathrm{Im}C_{eB}}{\Lambda^2}\bar{L}\sigma^{\mu\nu}H
e_RB_{\mu\nu}+h.c.\nonumber\\
\hat{O}_{eW}&=&i\frac{g\mathrm{Im}C_{eW}}{\Lambda^2}\bar{L}\sigma^{\mu\nu}\frac{\tau^i}{2}H
e_RW^i_{\mu\nu}+h.c.\nonumber\\
\hat{O}_{eH^3}&=&i\frac{\mathrm{Im}C_{eH^3}}{\Lambda^2}\bar{L}He_RH^\dagger
H+h.c.
\end{eqnarray}
The first-two operators are dipole-like while the third operator is
included as well because it mixes with the first two via electroweak
renormalization. Only $\hat{O}_{eW}$ contributes to $D$, as it is the only one containing a $W$ field. After electroweak symmetry-breaking, one finds
\begin{equation}
D=-\frac{4\sqrt{2}g_A^2}{g_V^2+3g_A^2}(\frac{m_e}{v})(\frac{v}{\Lambda})^2\mathrm{Im}C_{eW}\ \ \ .
\end{equation}
Note the presence of the $m_e/v$ suppression due to the existence of a
derivative in the operator $\hat{O}_{eW}$. The current upper bound on
the neutron $D$-coefficient implies
$(v/\Lambda)^2|\mathrm{Im}C_{eW}|<1\times 10^2$.

The same set of operators also induces an electron EDM, given by
\begin{equation}
d_e=-\frac{\sqrt{2}e}{v}(\frac{v}{\Lambda})^2(\mathrm{Im}C_{eB}-\mathrm{Im}C_{eW})
\end{equation}
The current upper bound on $d_e$ \cite{Baron:2013eja} implies
$(v/\Lambda)^2|\mathrm{Im}C_{eB}-\mathrm{Im}C_{eW}|<7.7\times10^{-13}$.

At first glance, it seems that one could simply choose
$\mathrm{Im}C_{eB}=\mathrm{Im}C_{eW}$ at low energy to avoid the EDM
constraint. We want to argue that, however, this choice is highly
unnatural because the operators in Eq. \eqref{eq:op} mix under
electroweak renormalization as
\begin{equation}
\frac{d\Theta}{d\ln\mu}=\left(\begin{array}{ccc}
\frac{151g'^{2}-27g^{2}}{192\pi^{2}} & -\frac{3gg'}{64\pi^{2}} & 0\\
-\frac{gg'}{16\pi^{2}} & \frac{-11g^{2}+3g'^{2}}{192\pi^{2}} & 0\\
-\frac{3g'(g^{2}-3g'^{2})}{16\pi^{2}} &
-\frac{9g(g^{2}-g'^{2})}{32\pi^{2}} &
-\frac{3(9g^{2}+7g'^{2})}{64\pi^{2}}\end{array}\right)\Theta
\end{equation}
where $\Theta=\left(\begin{array}{ccc} g'\mathrm{Im}C_{eB} &
g\mathrm{Im}C_{eW} & \mathrm{Im}C_{eH^3}\end{array}\right)^{T}$.
Numerically, if we assume that the bounds on the $D$-coefficient is
marginally satisfied at $\mu=m_W$ (i.e.
$(v/\Lambda)^2|\mathrm{Im}C_{eW}|=1\times 10^{2}$), then after the
electroweak renormalization we find that
$(v/\Lambda)^2|\mathrm{Im}C_{eB}-\mathrm{Im}C_{eW}|\approx 4.0$ at
$\mu=10$ TeV. However, this number has to be fine-tuned to a
precision level of $2\times10^{-11}\%$ in order to satisfy the EDM
bound at low energy, and therefore it is obviously not natural. The
dipole-like operators in the quark sector suffer from the same
problem. We conclude that, in the absence of RH neutrinos, EDM
constraints imply that the existence of an observable
$D$-coefficient is highly unlikely.


\section{\label{sec:conclusion}Conclusion}

If neutrinos are Dirac particles, implying the existence of light
$\nu_R$ in nature, then present limits on the $D$-coefficient
indicate that the mass scale of any associated PCTV interactions may
be quite significant: $\Lambda/c \gsim 1$ TeV, where $c$ denotes a
$d=6$ operator Wilson coefficient. The corresponding reach
associated with limits on the PCTV triple correlation in polarized
$^{19}$Ne decay are somewhat weaker, but nevertheless quite
interesting. The observation of a non-zero effect in a next
generation experiment with either the neutron or nuclei is not
precluded by constraints from EDM search null results. On the other
hand, the LHC results for $pp\to e+X+ MET$ present a greater
challenge, implying  at least an order of magnitude improvement in
neutron decay PCTV correlation sensitivity would be needed for
discovery of a non-zero $D$-coefficient. Should such an observation
occur, resolving the tension between a non-zero PCTV correlation
measurement and neutrino mass naturalness considerations would
provide an interesting challenge for model building.


\section*{Acknowledgements}
The authors wish to thank Jordy de Vries and Barry Holstein for many useful
discussions. This work was supported in part under U.S. Department of Energy contract
number DE-SC0011095 (MJRM and CYS) and NSF grant number is PHY-1205896 (BEM).

\bibliographystyle{prsty}
\bibliography{CYSbib}

\end{document}